\documentclass[prl,aps,showpacs,twocolumn,superscriptaddress,floatfix]{revtex4}

\usepackage{graphicx}
\usepackage{epsfig}
\usepackage{amsfonts}
\usepackage{amsmath,amssymb}
\usepackage{bm}

\begin{document}

\title{Microscopic calculations of isospin-breaking corrections to superallowed $\beta$-decay}

\author{W. Satu{\l}a}
\affiliation{Institute of Theoretical Physics, Faculty of Physics, University of Warsaw, \\ ul. Ho\.za
69, PL-00-681 Warsaw, Poland}

\author{J. Dobaczewski}
\affiliation{Institute of Theoretical Physics, Faculty of Physics, University of Warsaw, \\ ul. Ho\.za
69, PL-00-681 Warsaw, Poland}
\affiliation{Department of Physics, P.O. Box 35 (YFL),
University of Jyv\"askyl\"a, FI-40014  Jyv\"askyl\"a, Finland}

 \author{W. Nazarewicz}
\affiliation{Department of Physics and
  Astronomy, University of Tennessee, Knoxville, Tennessee 37996, USA}
\affiliation{Physics Division, Oak Ridge National Laboratory, P.O. Box
  2008, Oak Ridge, Tennessee 37831, USA}
\affiliation{Institute of Theoretical Physics, Faculty of Physics, University of Warsaw, \\ ul. Ho\.za
69, PL-00-681 Warsaw, Poland}

 \author{M. Rafalski}
\affiliation{Institute of Theoretical Physics, Faculty of Physics, University of Warsaw, \\ ul. Ho\.za
69, PL-00-681 Warsaw, Poland}

\date{\today}

\begin{abstract}
The superallowed $\beta$-decay rates that provide stringent constraints on physics beyond the Standard Model of
particle physics are affected by nuclear structure effects through  isospin-breaking corrections.
The  self-consistent isospin- and angular-momentum-projected nuclear density functional theory is used for the first time to compute
those corrections for a number of Fermi transitions in nuclei  from $A=10$ to $A=74$. The resulting leading element of the CKM matrix, $|V_{\text{ud}}|= 0.97447(23)$, agrees well with the recent result of Towner and Hardy [Phys. Rev. C {\bf 77}, 025501 (2008)].
\end{abstract}

\maketitle

Nuclear $\beta$ decays provide us with the crucial information about the electroweak force and constraints on physics beyond the Standard Model \cite{[Sev06],[Ram08]}. Of particular importance are superallowed Fermi transitions between the $J^\pi=0^+$ members of an isospin multiplet that can be used to test the conserved vector current (CVC) hypothesis and provide the most restrictive test of the unitarity of the Cabibbo-Kobayashi-Maskawa (CKM) matrix. Under assumptions of zero energy transfer and pure   isospin, the transition matrix elements for superallowed $\beta$ decays  do not depend on nuclear structure.

For actual nuclei, however, small corrections to the Fermi matrix element
of  $J=0^+,T=1 \rightarrow J=0^+,T=1$ superallowed transitions must be applied
(see Refs.~\cite{[Tow94],[Har05c],[Tow08],[Mar06]}
and Refs.\ quoted therein):
\begin{equation}\label{MFa}
|M_F^{(\pm )}|^2 = 2 (1+\delta_{\rm R}^\prime)(1+\delta_{NS} -\delta_C),
\end{equation}
where $\delta_C$ is a nuclear-structure-dependent isospin-breaking correction, and
$\delta_{\rm R}^\prime$  and $\delta_{NS}$ are radiative
corrections. The corrected product of statistical rate function $f$ and
partial half-life $t$ can be written as
\begin{equation}\label{ft}
      ft = \frac{{\cal F}t}{(1+\delta_{\rm R}^\prime)(1+\delta_{NS} -\delta_C)}
\end{equation}
with ${\cal F}t$ being nucleus independent.

In spite of theoretical uncertainties in evaluation of radiative and
isospin-breaking corrections, the superallowed $\beta$-decays
provide a stringent test of the CVC hypothesis. In turn, it is also the most precise
source of information on the leading element $V_{\text{ud}}$ of the CKM
matrix~\cite{[Tow08],[Nak10]}.
Indeed, with the CVC
hypothesis confirmed, $V_{\text{ud}}$ can be extracted from the data  by averaging over 13 precisely measured superallowed  $\beta$ transitions spreading over a broad range of nuclei from $A=10$ to $A=74$~\cite{[Tow08]}.

The main focus of this work is isospin-breaking corrections
$\delta_C$. This topic has been a subject of numerous theoretical
studies using different techniques
\cite{[Tow08],[Dam69],[Orm95a],[Sag96],[Lia09],[Mil09a],[Aue09]}. The
standard in this field has been set by Hardy and Towner (HT) \cite{[Tow77],[Har05c],[Tow08]} who
employed the nuclear shell model (SM) to account for configuration mixing and
the  mean-field approach to describe the radial mismatch of proton and neutron
single-particle (s.p.) wave functions.
Our  approach to  $\delta_C$ is based on the self-consistent
isospin- and angular-momentum projected nuclear density functional
theory (DFT) \cite{[Sat09sa],[Sat10s]}. This  framework can
simultaneously describe  various effects that profoundly impact
matrix elements of the Fermi decay; namely, symmetry breaking,
configuration mixing,  and long-range Coulomb polarization. It should
also be noted that our method is quantum-mechanically consistent (see
discussion in Ref.~\cite{[Mil09a]}) and contains no adjustable free
parameters.

The isospin- and angular-momentum projected DFT approach is based on
self-consistent states $|\varphi \rangle$
which, in general,  violate both rotational and isospin symmetries.
While the rotational invariance is broken spontaneously \cite{[Fra00],[Sat05]}, the isospin symmetry is
broken  both spontaneously (on DFT level) and directly by
the Coulomb force. Consequently,
the theoretical strategy  is to restore the rotational invariance,
remove  the spurious isospin mixing present in the DFT wave function, and retain only the physical isospin mixing
caused by the Coulomb interaction.
This is  achieved by the rediagonalization of the entire Hamiltonian,
consisting the isospin-invariant kinetic energy and nuclear interaction (Skyrme) terms, and
isospin-breaking Coulomb force,
in a good-angular-momentum and good-isospin basis
\begin{equation}\label{ITbasis}
|\varphi ;\, IMK;\, TT_z\rangle =   {\cal N}
\hat P^T_{T_z T_z} \hat P^I_{MK} |\varphi \rangle ,
\end{equation}
where $\hat P^T_{T_z T_z}$ and $\hat P^I_{MK}$ stand for the isospin
and angular-momentum projection operators and ${\cal N}$ is the normalization factor.
In the current version of the model,  nuclear isospin-breaking  interactions   and pairing have been disregarded.

The set of states (\ref{ITbasis}) is, in general, overcomplete because
the $K$ quantum number is not conserved. This difficulty is overcome
by selecting first the subset of linearly independent states
(collective space), which
is spanned, for each $I$ and $T$, by the  natural states
$|\varphi;\, IM;\, TT_z\rangle^{(i)}$ that are eigenstates of the overlap matrix \cite{[Dob09h],[Zdu07sa]}.
Diagonalization  of the Hamiltonian
in the collective space yields the eigenfunctions:
\begin{equation}\label{KTmix}
|n; \,\varphi ; \,
IM; \, T_z\rangle =  \sum_{i,T\geq |T_z|}
   a^{(n;\varphi)}_{iIT} |\varphi;\, IM; TT_z\rangle^{(i)} ,
\end{equation}
where the index $n$  labels eigenstates in ascending order
according to their energies while
$I$, $M$, and $T_z=(N-Z)/2$ are strictly
conserved. By construction, vectors (\ref{KTmix}) are free from spurious
isospin mixing. Moreover, since projection is applied to self-consistent DFT solution, a subtle interplay between the Coulomb polarization (that tends to make the proton and neutron wave functions different)
and the short-range nuclear attraction
(acting in exactly the opposite way) is properly taken into account.
As discussed in Refs.~\cite{[Sat09sa],[Sat10s],[Sat10a],[Sat10b]}, direct inclusion of monopole polarization effect, which is crucial for
 evaluation of isospin mixing in open-shell heavy nuclei, excludes all core-based models thus leaving us  with essentially one choice:  the nuclear
DFT. Recent experimental
data on  isospin impurities deduced in $^{80}$Zr from the giant dipole
resonance $\gamma$-decay studies \cite{[Cam10a]} are consistent with the magnitude
of isospin mixing  calculated with isospin-projected DFT \cite{[Sat10b]}, and this is very encouraging.

As demonstrated in Ref.~\cite{[Sat10a]}, in odd-odd $N=Z$ nuclei, the
isospin projection alone is not sufficient and a simultaneous
angular-momentum projection is a must. Unfortunately, this leads to
the appearance of singularities in the energy
kernels~\cite{[Sat10b]}, thus  preventing us from using modern  Skyrme
energy density functionals (EDFs) as none of them is usable, whereas
those depending on integer powers of the density, which are
regularizable \cite{[Lac09]}, are not yet developed. Hence, at
present, the only practical option is to use the Hamiltonian-driven EDFs
which, for  Skyrme-type functionals, leaves only one option: the
density-independent  SV parametrization \cite{[Bei75s]}  supplemented
by  tensor terms.

The unusual form of  SV   impacts negatively  its
overall spectroscopic quality by impairing such key properties as the
symmetry energy~\cite{[Sat10b]}, level density, and level ordering.
These deficiencies  affect the calculated isospin mixing. For
instance, for the case of $^{80}$Zr discussed above,  SV  yields the
isospin mixing 2.8\%, i.e.,  smaller  than the mean isospin
mixing 4.4\% averaged over nine commonly used  Skyrme EDFs,
see  Fig.~1 of~Ref.~\cite{[Sat10a]}. Of course, for the description
of $\delta_C$, of importance is not the absolute magnitude of isospin mixing but
its difference  between parent and daughter states \cite{[Aue09]}.
The  lack of reasonable EDF is, admittedly, the weakest point of our
current calculations; nevertheless, no significant improvement of this
aspect can be expected in the near future.

The $0^+  \rightarrow 0^+$ Fermi $\beta$-decay proceeds between the
$| I=0, T\approx 1, T_z = \pm 1 \rangle$ ground state (g.s.) of the
even-even nucleus and its isospin-analogue partner $|I=0, T\approx 1,
T_z = 0 \rangle$ in the $N=Z$ odd-odd nucleus. While the DFT state
representing the even-even nucleus is  unambiguously defined, the DFT
state used to compute the $N=Z$ wave function is the so-called
anti-aligned configuration $|\varphi \rangle \equiv |\bar \nu \otimes
\pi \rangle$ (or  $| \nu \otimes \bar \pi \rangle$), selected by
placing the odd neutron and the odd proton in the lowest available
time-reversed (or signature-reversed) s.p.\ orbits. The
anti-aligned  configurations  manifestly break the isospin symmetry
but they provide a way to reach the $|T\approx 1, I=0\rangle$  states
in odd-odd $N=Z$ nuclei \cite{[Sat10s]}. This situation creates
additional technical problems.  The anti-aligned configurations
appear to be very difficult to converge in the symmetry-unrestricted
DFT calculations. This can be traced back to time-odd components of
the EDF. In fact, only in a few cases were we able to obtain
symmetry-unrestricted self-consistent solutions. This forced us to
impose the signature-symmetry on  other DFT wave functions, which implied
a specific s.p.\ angular-momentum alignment pattern \cite{[Sch10a]}.

The calculations presented here were done using the  DFT solver HFODD (v2.48q)
\cite{[Dob09h]}, which
includes both the angular-momentum and isospin projection.
The calculated values of $\delta_C$ depend on the basis size.
In order to obtain converged result for $\delta_C$
with respect to basis truncation, we use
10 oscillator shells for $A< 40$ nuclei,
12 oscillator shells for $40\leq A< 62$ nuclei,
and 14 oscillator shells for $A\geq 62$ nuclei.
The  resulting systematic errors due the basis cut-off
do  not exceed $\sim$10\%.

The equilibrium quadrupole deformations ($\beta_2,\gamma$) of the anti-aligned configurations in odd-odd
nuclei are, in most cases, very close to those obtained for even-even
isobaric analogs. Typical differences do not exceed  $\Delta\beta_2 \approx 0.005$
and  $\Delta\gamma \approx 1^\circ$ except for nearly spherical systems
$A=14$ and $A=42$,  where the concept of static deformation is
ill-defined, and for $A=10$ and $A=18$ pairs where odd-odd and even-even partners have fairly different shapes. As we shall see below, such deformation difference results in large values of $\delta_C$.

All studied odd-odd nuclei, except for $A=14$, 38, and 42, are
deformed; thus, to carry out projections, we could use for them
the unique lowest anti-aligned DFT states. Also for $A=14$ and 38,
unique configurations based on the 1$p_{1/2}$ and 2$s_{1/2}$ subshells
were used.
A different approach was used to compute $\delta_C$ in near-spherical $A=42$
nuclei. In $^{42}$Sc,   four possible anti-aligned DFT configurations
built on the s.p.\ orbits originating from the spherical 1$f_{7/2}$
subshells can be formed, and the corresponding DFT states
differ slightly due to configuration-dependent polarizations
\cite{[Sat10a]}. Consequently, to evaluate $\delta_C$ for  $A=42$  we
took an arithmetic mean over the values calculated for all
anti-aligned configurations.

The unusually large correction $\delta_C\approx 10$\% has been calculated
for $A=38$ nuclei. Most likely,
this is a consequence of incorrect  shell structure
predicted with  SV. Specifically, as a result of incorrect balance
between the spin-orbit and tensor terms in SV, the $2s_{1/2}$ subshell
is shifted up to the Fermi surface. This state is
more sensitive to time-odd polarizations
than other s.p.\ states around $^{40}$Ca core, see
Table I in Ref.~\cite{[Zal08s]}. Consequently,
the $^{38}$K$\rightarrow$$^{38}$Ar transition has been excluded
from our calculation of  $V_{\text{ud}}$.

\begin{figure}
\includegraphics[angle=0,width=0.9\columnwidth,clip]{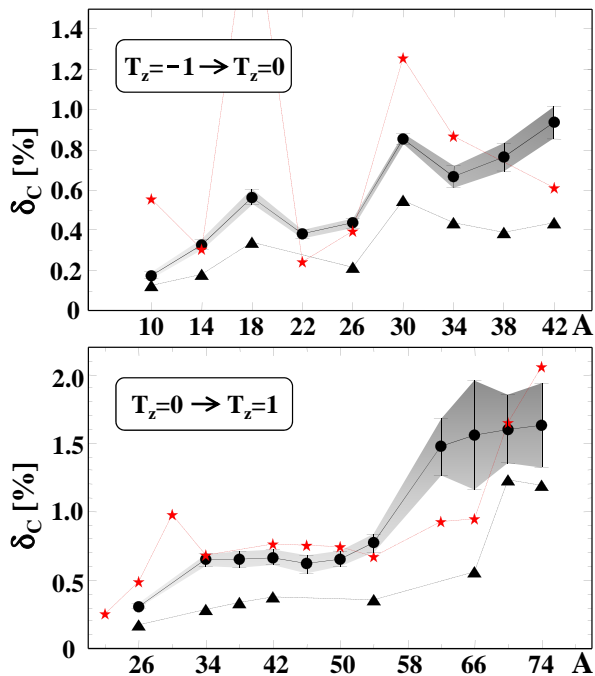}
\caption[T]{\label{fig4} (Color online)
Calculated isospin-mixing corrections $\delta_C$ for $T_z= -1
\rightarrow T_z =0 $ (top) and $T_z= 0 \rightarrow T_z =1 $ (bottom).
Our adopted values  (stars) are compared to values of
Ref.~\cite{[Tow08]} (dots; including errors) and Ref.~\cite{[Lia09]}
(triangles).
}
\end{figure}
The adopted values of $\delta_C$ are shown in Fig.~\ref{fig4} for
the $T_z=-1 \rightarrow T_z =0$
transitions  in light systems (upper panel) and
for  $T_z= 0 \rightarrow T_z = 1$ transitions pertaining
to heavier systems (lower panel). It is instructive to compare  our results to those of Refs.~\cite{[Tow08]} and \cite{[Lia09]}.
In light nuclei, the calculated $\delta_C$ are sensitive
to the local shell structure.  Indeed, although our values of  $\delta_C$ show  roughly the same trend as
those of Ref.~\cite{[Tow08]}, the individual
values differ. The reason can be traced
back to the poor spectroscopic quality of SV,  which manifests itself much
stronger in light than in heavier nuclei  due to the  low s.p.\ level density.  Let us also remind  that the
equilibrium minima in parent and daughter nuclei with $A=10$ and $A=18$  differ, and this results in increased $\delta_C$ values. As verified by DFT calculations using other EDFs, and also findings of Ref.~\cite{[Lia09]}, with higher level density in heavier nuclei the detailed shell structure seems to play a lesser role. This indicates
that gross features of configuration mixing in heavier nuclei
associated with long-range time-even (shape) correlations are less
dependent on a EDF parametrization  and may be
relatively well captured  by  SV. The calculated
values of  $\delta_C$  for heavier nuclei are indeed  quite
consistent with the HT results \cite{[Tow08]}, with the exception
of $A=62$ and 66.

\begin{table}
\caption[A]{Experimental $ft$-values (in sec);  $\delta_C$ values
adopted in this work (in \%);  calculated ${\cal F}t$-values (in sec);
empirical corrections  (\ref{expdelt}) (in \%),   and individual contributions
to $\chi^2$ used in the CL test. } \label{tab2}
\begin{tabular}{lrrrcrr}
\hline\hline
Parent    &   {$ft$~~~~~~}       &   {$\delta_C$~~~~}  &  {${\cal F}t$~~~~}  &
$\delta_C^{{\rm (EXP)}}$  &  $\chi^2_i$  \\
\hline
$T_z=-1:$ &              &           &              &           &     \\
$^{10}$C  &  3041.7(43)  &  0.559(56)&  3064.8(48)  &  0.39(14) &  1.3\\
$^{14}$O  &  3042.3(11)  &  0.303(30)&  3072.3(21)  &  0.38(06) &  1.5\\
$^{22}$Mg &  3052.0(70)  &  0.243(24)&  3082.2(71)  &  0.64(23) &  3.0\\
$^{34}$Ar &  3052.7(82)  &  0.865(87)&  3063.5(87)  &  0.65(27) &  0.6\\
\hline
$T_z=0: $ &             &            &               &           &     \\
$^{26}$Al &  3036.9(09) &   0.494(49)&   3066.7(20)  &  0.39(04) &  6.8\\
$^{34}$Cl &  3049.4(11) &   0.679(68)&   3069.8(26)  &  0.67(05) &  0.0\\
$^{42}$Sc &  3047.6(12) &   0.767(77)&   3069.2(31)  &  0.74(06) &  0.1\\
$^{46}$V  &  3049.5(08) &   0.759(76)&   3069.0(30)  &  0.73(06) &  0.3\\
$^{50}$Mn &  3048.4(07) &   0.740(74)&   3068.3(31)  &  0.69(07) &  0.7\\
$^{54}$Co &  3050.8(10) &   0.671(67)&   3073.0(32)  &  0.77(08) &  1.5\\
$^{62}$Ga &  3074.1(11) &   0.925(93)&   3088.7(41)  &  1.52(09) & 41.0\\
$^{74}$Rb &  3084.9(77) &   2.06(21) &   3064(11)    &  1.88(27) &  0.4\\
\hline\hline
\end{tabular}
\end{table}

The predicted isospin-breaking corrections  are listed in Table~\ref{tab2}.
All other ingredients needed to compute  ${\cal F}t$-values from
Eq.~(\ref{ft}), including empirical $ft$-values and radiative
corrections $\delta_{\rm R}^\prime$ and $\delta_{NS}$, were taken
from the most recent compilation~\cite{[Tow10]}.
In the error budget  of ${\cal F}t$ in Table~\ref{tab2}, apart from errors of $ft$
and radiative corrections,  we include 10\%
systematic uncertainty in the calculated $\delta_C$ due to basis truncation.
The average value $\overline{{\cal F}t} = 3070.4(9)$\,s was obtained
using Gaussian-distribution-weighted formula to conform with standards set by HT.
This leads to $|V_{\text{ud}}| = 0.97447(23)$ which
coincides with both the HT result  $|V_{\text{ud}}^{\text{(HT)}}| = 0.97418(26)$
\cite{[Tow08]} and a central value obtained from the neutron decay
$|V_{\text{ud}}^{(\nu )}| = 0.9746(19)$ \cite{[Nak10]}.
Combining the calculated $|V_{\text{ud}}|$ with
the values of $|V_{\text{us}}| = 0.2252(9)$ and  $|V_{\text{ub}}| = 0.00389(44)$
provided in Ref.~\cite{[Nak10]}, we obtain $|V_{\text{ud}}|^2 +  |V_{\text{us}}|^2  + |V_{\text{ub}}|^2 =  1.00031(61)$,
which implies that unitarity of the CKM matrix is satisfied
with precision of 0.1\%.


\begin{figure}[t]
\begin{center}
\includegraphics[angle=0,width=0.8\columnwidth,clip]{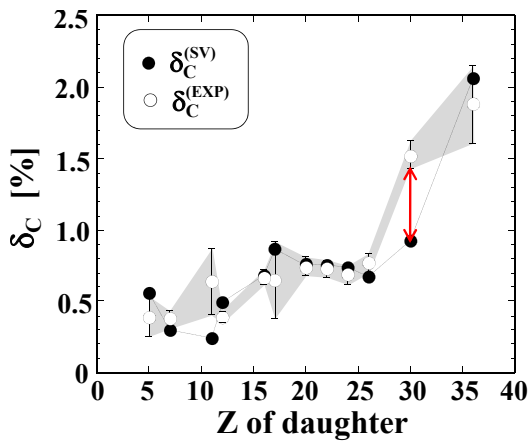}
\caption[T]{\label{fig6}(Color online)
Calculated (black dots) and empirical (white dots, with error bars) values of $\delta_C$ as function of
proton number in the daughter nuclei. Vertical arrow marks the values for  $A=62$. See text for details.
} \end{center}\end{figure}

While our value of $|V_{\text{ud}}|$ is consistent
with both HT and neutron-decay results, a
question arises about its  confidence level, especially in light of
poor spectroscopic properties of SV. To this end,
we carry out  the confidence-level (CL) test
proposed recently in  Ref.~\cite{[Tow10]}
using variant including uncertainties on experiment, $\delta_{\rm R}^\prime$, and
$\delta_{NS}$. The  test
is based on the assumption that the CVC hypothesis is valid
to at least $\pm 0.03$\% precision, implying that a set of
structure-dependent corrections should produce a statistically consistent
set of ${\cal F}t$ values. Since  only one set of
calculated  $\delta_{NS}$ corrections exists~\cite{[Tow94]}, ``empirical"
isospin-symmetry-breaking corrections can thus be defined by
\begin{equation}\label{expdelt}
\delta_C^{{\rm (EXP)}}  = 1 + \delta_{NS}
- \frac{\overline{{\cal F}t}}{ft(1+\delta_{\rm R}^\prime)},
\end{equation}
and they are tabulated in Table~\ref{tab2}.
The CL can be assessed by minimizing the  root mean square deviation between predicted and empirical values of $\delta_C$  with respect to $\overline{{\cal F}t}$ in Eq.~(\ref{expdelt}).
The final result corresponding to $\overline{{\cal F}t} = 3070.0$\,s is shown in Fig.~\ref{fig6}.
Individual contributions to  $\chi^2$ are also displayed in  Table~\ref{tab2}. Our value of reduced $\chi^2$ (per degree of freedom; in our case $n_d=11$) is  5.2.
This  is considerably higher than the values reported in Ref.~\cite{[Tow10]}
for the Damgaard model~\cite{[Dam69],[Tow77]} (1.7),
SM with Woods-Saxon radial wave functions \cite{[Tow08]} (0.4),
SM with Hartree-Fock (HF) radial wave functions \cite{[Orm95a],[Har05c]} (2.2), and relativistic
Hartree plus RPA model of~\cite{[Lia09]} (2.1).
The low CL of our model results primarily from the single point at  $A=62$.

In summary,  the state-of-the-art isospin- and angular-momentum-projected DFT
calculations have been performed to compute the isospin-breaking
corrections to $0^+ \rightarrow 0^+$ Fermi superallowed $\beta$-decays.
Our results for  $\bar{\cal F}t = 3070.4(9)$\,s and  $|V_{\text{ud}}| =
0.97447(23)$ were found to be  consistent with the recent HT value
\cite{[Tow08]}. While the CL of our $\delta_C$ values is low,
primarily due to a poor spectroscopic quality of the EDF  used, our
framework contains no adjustable parameters and is capable of
describing microscopically all elements of physics impacting
$\delta_C$. The results presented in this paper should thus be
considered as a microscopic benchmark relative to which the further
improvements (e.g., regularizable EDF and/or
inclusion of pairing) will be assessed.

This work was supported in part by the Polish Ministry of Science
under Contract Nos.~N~N202~328234 and N~N202~239037, Academy of Finland and
University of Jyv\"askyl\"a within the FIDIPRO programme, and by the Office of
Nuclear Physics,  U.S. Department of Energy under Contract Nos.
DE-FG02-96ER40963 (University of Tennessee) and
DE-FC02-09ER41583  (UNEDF SciDAC Collaboration).
We acknowledge the CSC - IT Center for Science Ltd, Finland for the
allocation of computational resources.


\end{document}